\documentstyle[graphicx,aps,twocolumn]{revtex}

\newcommand{\myscalebox}[1]{\scalebox{0.4}[0.45]{#1}}
\newcommand{\captionA}
{Distribution of overlaps $P(|q|)$  for ground
states of 3d $\pm J$ Ising spin glass for $L=6,12$. Only for large
values of $q$ a difference is visible, so even for large systems there
is a finite probability of overlap $q=0$. The lines are guides for the 
eyes only.}

\newcommand{\captionB}
{Distribution $P(\delta q)$ for different system sizes $L=4,10,14$ 
where $\delta q=q_2-q_1$ and $q_1\le q_2\le q_3$ are
triplets of absolute values of overlaps from 
independent triplets of ground states.
Only triplets with $q_3 \in [0.5,0.6]$ are used. For a infinite ultrametric
system $\delta q=0$ holds. With increasing system size
the distributions get closer to $q=0$ indicating the increasing 
ultrametricity
of the ground states. The lines are guides for the 
eyes only.}

\newcommand{\captionC}
{Average value of $\delta q$ as function of system size $L$ 
where $\delta q=q_2-q_1$ and $q_1\le q_2\le q_3$ are
triplets of absolute values of 
overlaps from independent triplets of ground states. 
Only triplets with $q_3 \in [0.5,0.6]$ are used.With increasing system size
the average decreases, which indicates the increasing ultrametricity
of the ground states. The straight line represents the
function $\langle \delta q \rangle(L)=0.235 \times L^{-0.255}$.}

\newcommand{\captionD}
{Distribution $P_{2-fix}(q)$ for different system sizes $L=4,8,12$ 
where $q\in\{q_1,q_2,q_3\}$ and $q_1\le q_2\le q_3$ are
triplets of overlaps from independent triplets of ground states. 
Only $q$-values of triplets are used where the two other overlaps 
are within the
interval $[0.5,0.6]$. Then for a infinite ultrametric system 
$q>0.5$ holds, while for a metric system just $q>0$ must hold. The small inset
shows the part $q\in[0.0,0.5]$ for $L=4,6,8,10,12$ (from left to right). 
With increasing system
size the fraction of the distribution below $0.5$ shrinks, so the systems
become more and more ultrametric. The lines are guides for the 
eyes only.}

\newcommand{\captionE}
{Integrated value $I_L=\int_{-1}^{0.5}P_{2-fix}(q)(q-0.5)^2\,dq$
$+\int_{0.89}^1 P_{2-fix}(q)(q-0.89)^2\,dq$ 
as function of system size $L$ 
where $q\in\{q_1,q_2,q_3\}$ and $q_1\le q_2\le q_3$ are
triples of overlaps from independent triples of ground states.
Only $q$-values of triples where the two other overlaps are within the
interval $[0.5,0.6]$ are used. 
With increasing system
size the fraction of the distribution outside $[0.5,0.89]$ decreases,  
so the systems
become more and more ultrametric. The straight line represents the
function $I(L)=1.2\times L^{-0.61}$.}

\newpage
\newcommand{\figA}{
\begin{figure}[t]
\begin{center}
\myscalebox{\includegraphics{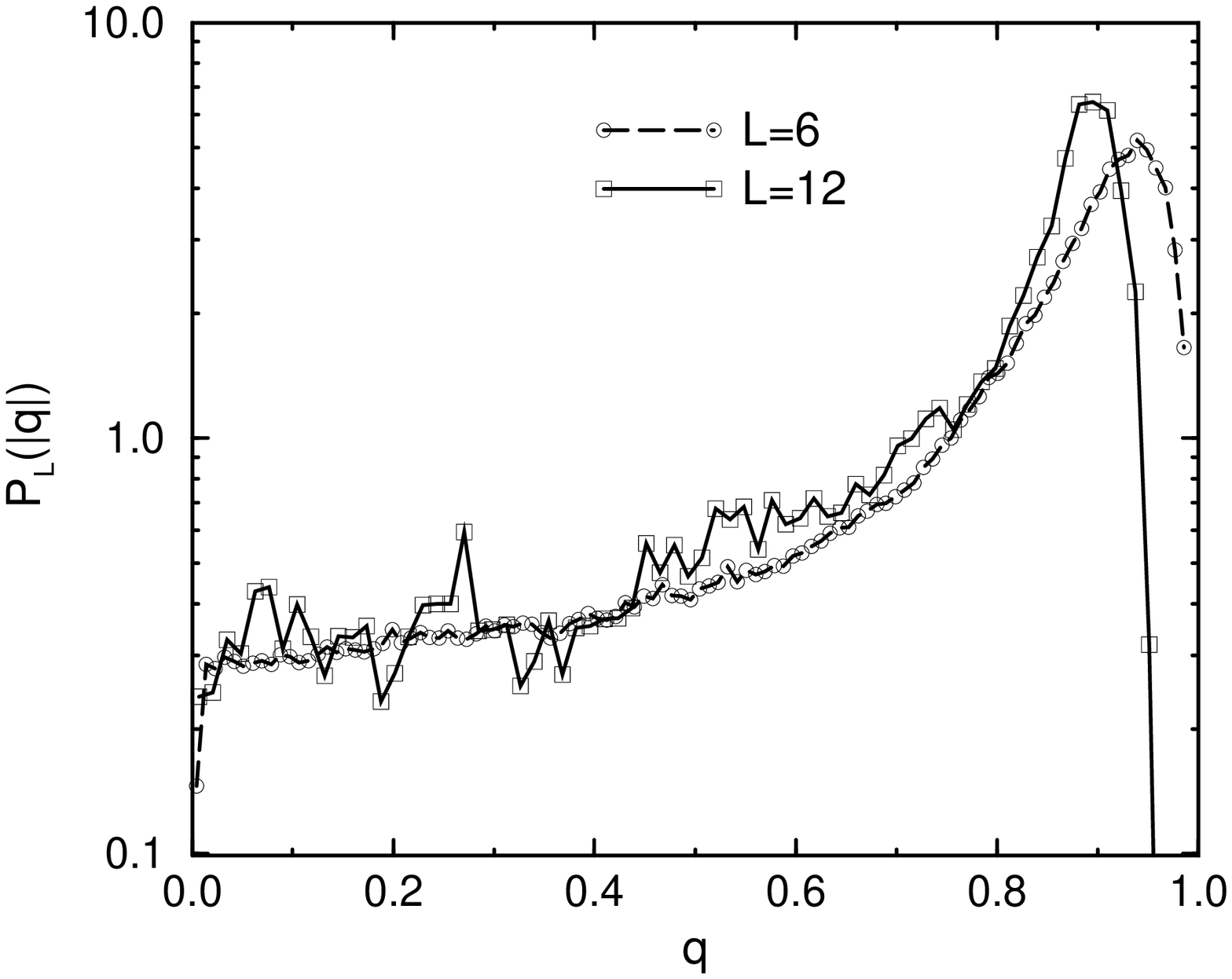}}
\end{center}
\caption{\captionA}
\label{fig_plq}
\end{figure}
}

\newcommand{\figB}{
\begin{figure}[t]
\begin{center}
\myscalebox{\includegraphics{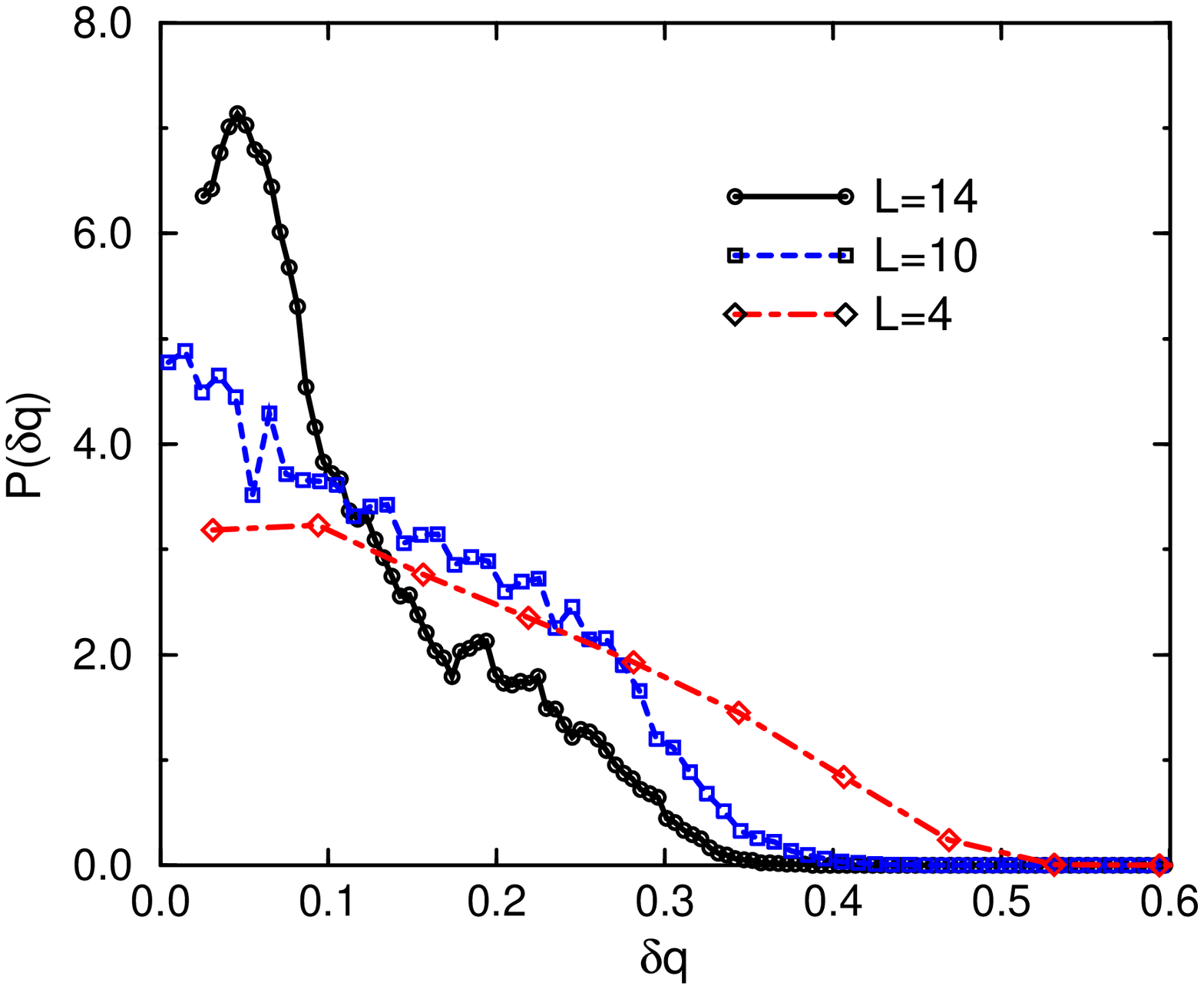}}
\end{center}
\caption{\captionB}
\label{fig_deltaq}
\end{figure}
}

\newcommand{\figC}{
\begin{figure}[t]
\begin{center}
\myscalebox{\includegraphics{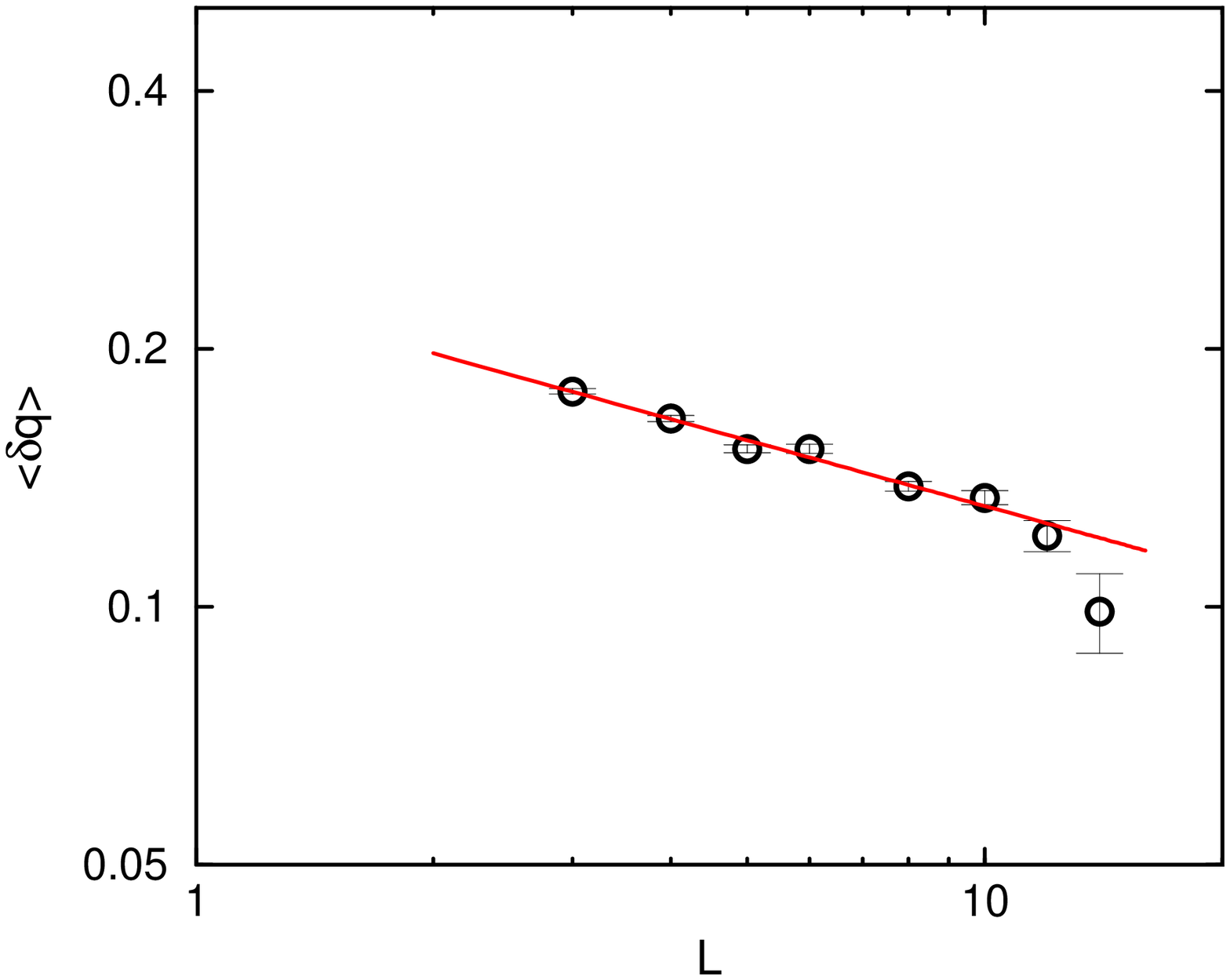}}
\end{center}
\caption{\captionC}
\label{fig_deltaq_mw}
\end{figure}
}

\newcommand{\figD}{
\begin{figure}[t]
\begin{center}
\myscalebox{\includegraphics{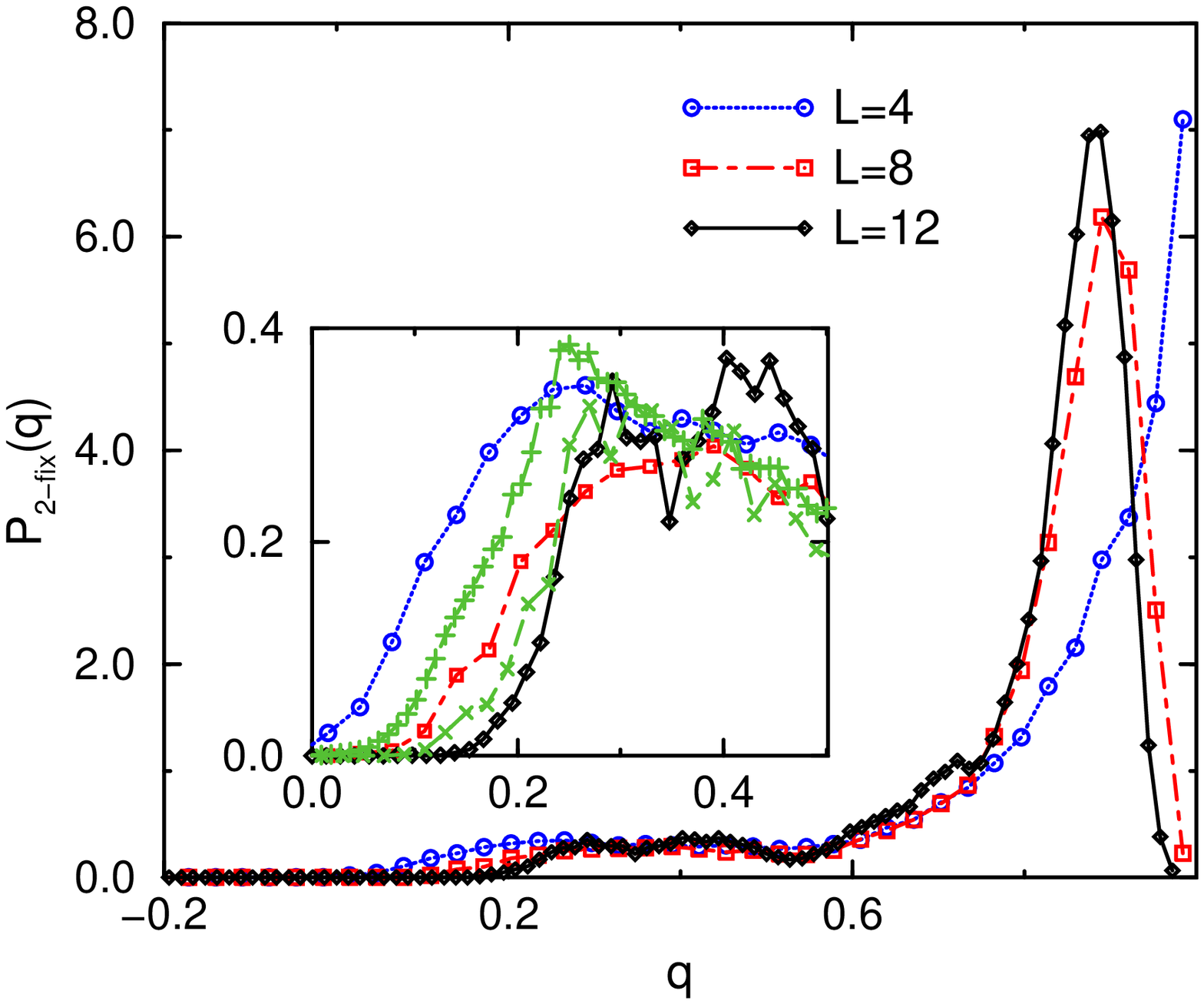}}
\end{center}
\caption{\captionD}
\label{fig_pq_2fix}
\end{figure}
}

\newcommand{\figE}{
\begin{figure}[t]
\begin{center}
\myscalebox{\includegraphics{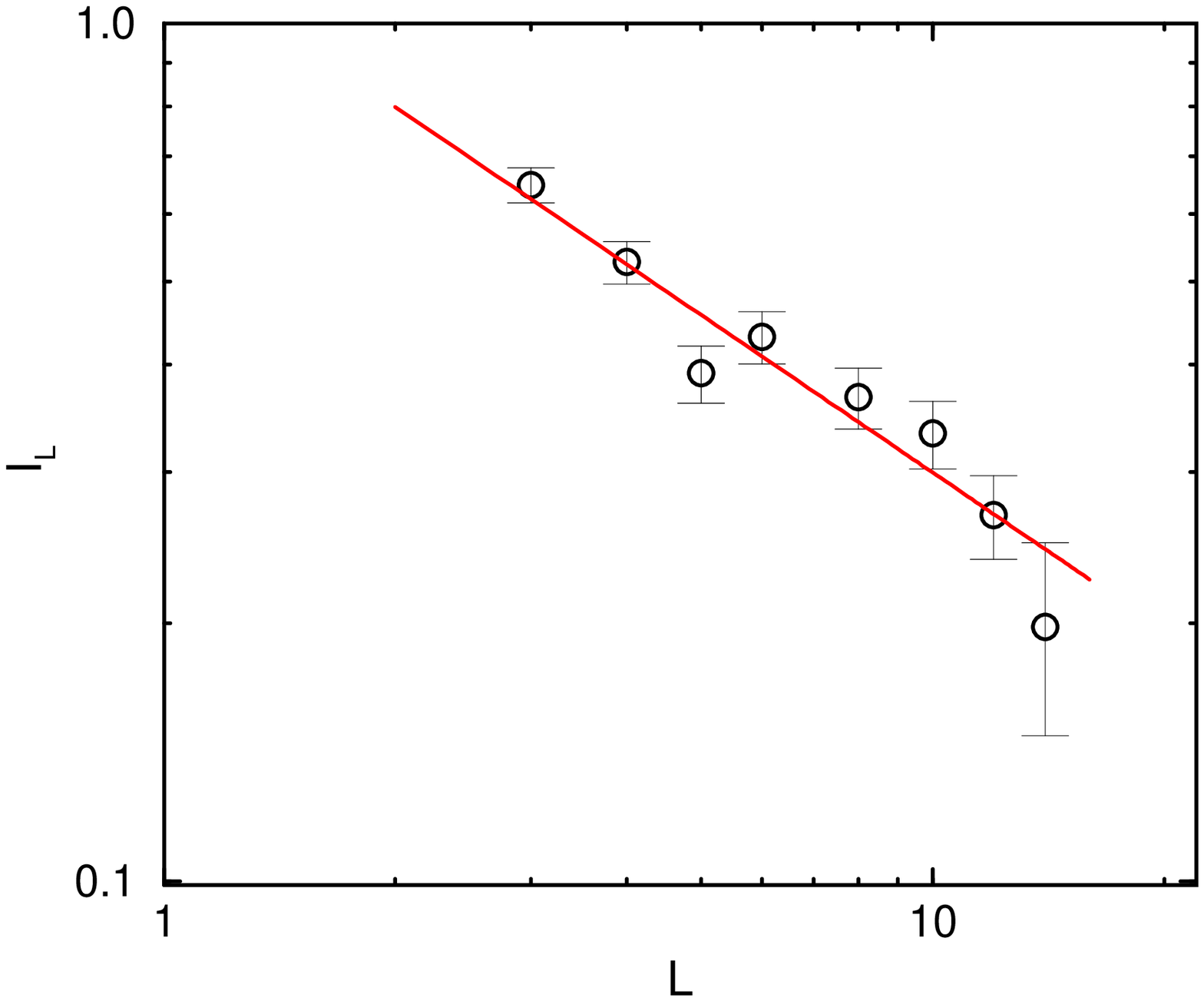}}
\end{center}
\caption{\captionE}
\label{fig_pq_2fix_int}
\end{figure}
}

\begin{document}
\title{Are Ground States of 3d $\pm J$ Spin Glasses ultrametric ?}

\author{Alexander K. Hartmann\\
{\small  hartmann@tphys.uni-heidelberg.de}\\
{\small  Institut f\"ur theoretische Physik, Philosophenweg 19, }\\
{\small 69120 Heidelberg, Germany}\\
{\small Tel. +49-6221-549449, Fax. +49-6221-549331}}

\date{\today}
\maketitle
\begin{abstract}
Ground states of 3d EA Ising spin glasses are calculated
for sizes up to $14^3$ using a combination 
of a genetic algorithm and Cluster-Exact
Approximation. Evidence for an ultrametric structure is found by
studying triplets of independent 
ground states where one or two values of the three
overlaps are fixed.

{\bf Keywords (PACS-codes)}: Spin glasses and other random models(75.10.Nr), 
Numerical simulation studies (75.40.Mg),
General mathematical systems (02.10.Jf). 
\end{abstract}


\paragraph*{Introduction}

The behavior of the Edwards-Anderson (EA) $\pm J$ Ising spin glass with short
range (i.e. realistic) interactions is still not well understood. 
Introduction to spin glasses can be found for example 
in \cite{binder86,fisher91}.
The EA Ising spin glass is a system of $N$ spins 
$\sigma_i = \pm 1$, described by the Hamiltonian

\begin{equation}
H \equiv - \sum_{\langle i,j\rangle} J_{ij} \sigma_i \sigma_j
\end{equation}

The sum $\langle i,j\rangle$ goes over nearest neighbors. In this letter we
consider 3d cubic systems with periodic boundary conditions,
$N=L^3$ spins and the exchange interactions (bonds) take
$J_{ij} = \pm 1$ with equal probability under the constraint
$\sum_{<i,j>} J_{ij} = 0$.

This work addresses the question whether the ground states of this system 
exhibit an
{\em ultrametric } structure: The distances $d^{\alpha\beta}$ between states 
do not only satisfy the triangular inequality
$d^{\alpha\beta} \le d^{\alpha\gamma} + d^{\gamma\beta}$
but the stronger ultrametric inequality
$d^{\alpha\beta} \le \max(d^{\alpha\gamma}, d^{\gamma\beta})$ as well.
For an
introduction to ultrametricity see \cite{rammal86}.
The mean field solution \cite{parisi2} of the infinite-dimensional
SK-model \cite{sherrington} shows an ultrametric state structure
\cite{mezard84,franz92}. Numerical work on the subject can be found in
\cite{parga84,bhatt86}. For finite dimensional
systems numerical evidence for ultrametricity 
at finite temperature but below the transition Temperature $T_G$ was
found in four dimensions \cite{cacciuto97,marinari98}. First attempts
to find ultrametricity in three dimensions by simulation at
finite temperature are given in \cite{sourlas84,caracciolo90}.

Here the first time ground states of realistic spin glasses 
are analyzed regarding their ultrametric structure. 
The ground state structure has
a strong influence on the overall behavior of a system:
an ultrametric ground state structure implies a complex free
energy landscape so that
no efficient variant of the usual cluster algorithms exists, i.e. that there
is a critical slowing down \cite{persky96}.

In this letter we show our results for the direct calculation of spin glass
ground states using a hybrid of genetic algorithms \cite{pal1,michal} 
and Cluster-Exact Approximation (CEA) \cite{alex2}. Because the computation of
 spin glass ground states belongs to the class of NP-hard problems, it is a 
tough computational task. Using this new
algorithms it is possible the first time to calculated ground states
of adequate size (up to $L=14$) and with sufficient statistics (especially
for the largest sizes). It is possible to calculate many strictly statistical
independent configurations (replicas). 
In contrast to Monte Carlo methods one does not
encounter ergodicity problems. It is importance to notice that no kind of 
temperature is involved in our method.

\paragraph*{Observables}

For a fixed realization $J=\{J_{ij}\}$ of the exchange interactions and two
replicas
$\{\sigma^{\alpha}_i\}, \{\sigma^{\beta}_i\}$, the overlap \cite{parisi2}
is defined as

\begin{equation}
q^{\alpha\beta} \equiv \frac{1}{N} \sum_i \sigma^{\alpha}_i \sigma^{\beta}_i 
\end{equation}

The ground state of a given realization is characterized by the probability
density $P_J(q)$. Averaging over the realizations $J$, denoted
by $[\,\cdot\,]_{av}$, results in ($Z$ = number of realizations)

\begin{equation}
P(q) \equiv [P_J(q)]_{av} = \frac{1}{Z} \sum_{J} P_J(q) \label{def_P_q}
\end{equation}

Because no external field is present the densities are symmetric:
$P_J(q) = P_J(-q)$ and $P(q) = P(-q)$. So we calculate only functions
$P_J(|q|)$ and $P(|q|)$.

The overlap measures the distance between two states. This can be
reflected by defining a distance function

\begin{equation}
d^{\alpha\beta} \equiv 0.5(1-q^{\alpha\beta})
\end{equation}

with $0\le d^{\alpha\beta} \le 1$. 
For three replicas $\alpha,\beta,\gamma$ the usual triangular inequality
reads
$d^{\alpha\beta} \le d^{\alpha\gamma} + d^{\gamma\beta}$.
Expressed in terms of $q$ it becomes

\begin{equation}
q^{\alpha\beta} \ge q^{\alpha\gamma} + q^{\gamma\beta}-1
\label{triangular_q}
\end{equation}

In an {\em ultrametric} space \cite{rammal86} 
the triangular inequality is replaced by a stronger one
$d^{\alpha\beta} \le \max(d^{\alpha\gamma}, d^{\gamma\beta})$
or equivalently

\begin{equation}
q^{\alpha\beta} \ge \min(q^{\alpha\gamma}, q^{\gamma\beta})
\label{ultra_q}
\end{equation}

An example of an ultrametric space is the set of leaves of a binary tree:
The distance between two leaves is defined by the number of edges on a path
between the leaves.
Let $q_1\le q_2 \le q_3$ be the overlaps $q^{\alpha\beta}$, 
$q^{\alpha\gamma}$, 
$q^{\gamma\beta}$ ordered according their sizes.
By writing the smallest overlap on the left side in Equation (\ref{ultra_q}), 
one realizes that two of the overlaps must be equal and 
the third may be larger or the same: $q_1 = q_2 \le q_3$

In a finite size system this relation may be violated.
We use two ways of determining whether ground states of
realistic spin glasses
become more and more ultrametric with increasing size $L$:

\begin{itemize}
\item
The difference 
\begin{equation}
\delta q\equiv q_2-q_1
\label{def_delta}
\end{equation}
 is calculated for all triplets. 
Because we want to exclude the influence of the absolute size
of the overlaps the third overlap is fixed: $q_3=q_{fix}$. In practice 
only overlap triples are used where $q_3 \in [q_{fix},q_{fix2}]$ holds
to obtain sufficient statistics . With
increasing size $L$ the distribution $P(\delta q)$ should tend to a 
Dirac delta function \cite{bhatt86}.
\item
If two overlaps are fixed ($q^{\alpha\gamma}=q^{\beta\gamma}=q_{fix}$,
in practice $q^{\alpha\gamma},q^{\beta\gamma}\in [q_{fix},q_{fix2}]$), 
equation (\ref{triangular_q}) implies $q\equiv q^{\alpha\beta}\ge 2q_{fix}-1$
while ultrametricity implies $q\ge q_{fix}$ which is stronger if $q_{fix}<1$
\cite{cacciuto97}. The
distribution $P_{2-fix}(q)$ of the third overlap is used to characterize
the ultrametricity of a system. 
The fraction of the distribution outside $[q_{fix},q_{EA}]$ 
\begin{eqnarray}
I_L & \equiv & \int_{-1}^{q_{fix}}P_{2-fix}(q)(q-q_{fix})^2\,dq \\\nonumber
& & + \int_{q_{EA}}^{1}P_{2-fix}(q)(q-q_{EA})^2\,dq
\end{eqnarray}
(see \cite{cacciuto97}) 
should vanish for $L\to\infty$ in an ultrametric system.
\end{itemize}

\paragraph*{Results}

We performed ground state calculations for sizes 
$L=$ 3, 4, 5, 6, 8, 10, 12, 14. 
For each size we used different parameter sets, 
which were determined in a way, that no decrease of the energy could be 
found by doubling the running time for some sample systems.
Using our parameters results roughly in an exponential increase of
the running time as function of $L$. 
One $L=14$ run needs typically 540 CPU-minutes on a 80MHz PPC601
processor
(70 CPU-minutes for $L=12$, $\ldots$, 0.2 CPU-seconds for $L=3$).
Details of the algorithm, simulation
parameters and more results will be given in \cite{alex3}.

For small system sizes $L=4,6$ we could compute for
200 randomly selected systems exact ground states with a Branch-and-Cut
method \cite{simone95,simone96} 
using a program which  generously was made available by the
group of M. J\"unger in Cologne. 
In ALL cases our method found the exact ground states as well! So we
are pretty sure that our algorithm computes true ground states or at least
states very close to true ground states even for larger systems.

We calculated from 136 realizations for $L=14$ up to 8900 realizations
for $L=3$. For each realization up to $40$ independent runs were made 
(up to $80$ for some large systems). Each run resulted in one 
configuration, which was
stored, if it exhibited the ground state energy. For $L=14$ this resulted
in an average of $n_{gs}=14.9$ states per realization having the lowest energy 
while for $L=3$ on average $n_{gs}=39.97$ states  were stored. 
This reduction of $n_{gs}$ 
means that with increasing system size true ground states are harder to
find, but not that the number of existing ground states is reduced:

\figA

In fig. \ref{fig_plq} the probability density of the overlap is
displayed for $L=6$ and $L=12$. Each realization enters the
sum with the same weight, even if different numbers of ground states
were available for the calculation of $P_J(|q|)$.
Only a small difference for high $q$-values can be observed. 
But especially for $|q|\le 0.8$ no significant
reduction in the probability is visible. So with increasing sizes the
width of the distribution remains finite \cite{alex_sg2}. It means that
realistic spin glasses have many arbitrary different 
ground states which are arranged in a complex  structure.

\figB

To investigate whether this structure is even ultrametric we have
calculated the quantity $\delta q$ (see equation (\ref{def_delta})) for
all possible triplets with $q_3 \in [0.5,0.6]$. To improve the statistics
we used the absolute value of all overlaps. The distribution $P(\delta q)$
is show in fig. \ref{fig_deltaq} for $L=4,10,14$. Each realization enters
the distribution with the same weight. With increasing system
size the distributions get closer to $q=0$, indicating that the systems become
more and more ultrametric. It is important to notice that the change in
$P(\delta q)$ is not due to a change of $P(q)$ itself as it could
not be excluded in
\cite{caracciolo90}. Since the shape of $P(q)$ does not change very much
with system size the effect on $P(\delta q$) is caused by the
increasing ultrametricity of the systems.

\figC

The same result is obtained by computing the
average values of $\delta q$ as function of the size $L$ (see fig. 
\ref{fig_deltaq_mw}). One gets a similar figure by computing the
variances of the distributions, but it is not shown here. 
The number of realizations
is to small too perform a reasonable fit of the form 
$\langle\delta q \rangle (L)=\langle\delta q \rangle_{\infty}+cL^{-\alpha}$. 
So no decision is  possible yet whether the average value converges 
to zero or not. We only provide the result of a fit
with $f_{\infty}$ set to zero. Then we get $c=0.235(2)$ and $\alpha=0.255(3)$.
Since the data has a small negative curvature the assumption 
$\langle\delta q \rangle_{\infty}=$ seems reasonable.

\figD

By fixing two of the three overlaps of a triplet we have another way of 
checking ultrametricity. We took all triplets where two arbitrary 
overlaps fell into
the interval $[0.5, 0.6]$ The resulting distributions $P_{2-fix}(q)$ 
of the
third remaining overlap is shown in fig. \ref{fig_pq_2fix} for $L=4,8,12$.
The triangular inequality gives $q>0$ while ultrametricity leads
to $q>0.5$.
The inset magnifies the values of $q\in[0.0,0.5]$ for $L=4,6,8,10,12$ (from 
left to right). The fraction of the distribution below $q=0.5$ shrinks
with increasing size. It is clearly visible that even the smallest
sizes are far away from the triangular bond $q>0$, but the system sizes
are too small to decide whether $q>0.5$ really holds for large systems.

\figE

 Fig. \ref{fig_pq_2fix_int} shows the distributions
integrated outside the interval $[q_{fix},q_{EA}]$
 as function of system size. We used $q_{EA}=0.89$ from the results
of \cite{alex_sg2}. The value $I_L$ decreases
with increasing size, but using a fit $I(L)=I_{\infty}+kL^{-\beta}$
no decision can be taken if it converges to zero. We only provide the result
of a fit with $I_{\infty}\equiv 0$, where we get 
$k=1.2(1)$ and $\beta=-0.61(7)$.

\paragraph*{Conclusion}

By the calculation of ground states using genetic Cluster-Exact Approximation
we find evidence for the existence of an ultrametric ground state
structure in short range $\pm J$ spin glasses. For more quantitative
statements the system sizes are too small, but it is clear that
the structure is more complex than simply metric. For treating larger systems
more elaborate algorithms or much faster computers must be available, 
because even for our results 32 PPC-601 processors where busy for 
more than three months 24 hours a day.

For spin glasses with Gaussian distribution
the ground state is not degenerate, but we 
expect the same behavior as for the $\pm J$ model
if one allows deviations of order one from the true ground state energy,

Concluding we believe that for realistic spin glasses
the scenario of an ultrametric organization
of the states  is more probable than a scenario with
a simple structure. 

\paragraph*{Acknowledgements}

We thank H. Horner and G. Reinelt for manifold support.
We are grateful to M. J\"unger, M. Diehl and T. Christof who made us available
a Branch-and-Cut Program for the exact calculation of spin glass
ground states of small systems.
We thank R. K\"uhn for critical reading the manuscript and for giving 
many helpful hints.
We took much benefit from discussions with
H. Kinzelbach, S. Kobe, H. Rieger, A.P. Young, 
N. Kawashima, N. Sourlas, J.-C. Angl\`{e}s d'Auriac, 
M. M\'{e}zard and R. Monasson.
We are grateful to the {\em Paderborn Center for Parallel Computing}
 for the allocation of computer time as well. This work was supported
by the Graduiertenkolleg ``Modellierung und Wissenschaftliches Rechnen in 
Mathematik und Naturwissenschaften'' at the
{\em In\-ter\-diszi\-pli\-n\"a\-res Zentrum f\"ur Wissenschaftliches Rechnen}
 in Heidelberg.

\newpage




\end{document}